\documentstyle[sprocl,epsfig,bezier]{article}
\newcommand{\lesssim}{\raisebox{0.3mm}{\em $\, <$} \hspace{-2.8mm}
\raisebox{-1.3mm}{\em $\sim \,$}}
\newcommand{\gtrsim}{\raisebox{0.3mm}{\em $\, >$} \hspace{-2.8mm}
\raisebox{-1.3mm}{\em $\sim \,$}}

\begin{document}

\title{Physics potential and present status of neutrino factories}

\author{OSAMU YASUDA}

\address{Department of Physics,
Tokyo Metropolitan University \\
Minami-Osawa, Hachioji, Tokyo 192-0397, Japan
\\E-mail: yasuda@phys.metro-u.ac.jp}


\maketitle\abstracts{
I briefly review the recent status of research on physics potential
of neutrino factories with emphasis on measurements of the CP phase.}

\section{Introduction}

The observation of atmospheric neutrinos (See, e.g.,
Ref.\cite{Kajita:2001mr})
and solar neutrinos (See, e.g., Ref.\cite{Bahcall:2000kh})
gives the information on the mass squared differences
and the mixings, which can be written in the three flavor
framework of neutrino oscillations as
($|\Delta m_{32}^2|$, $\theta_{23}$) and
($\Delta m_{21}^2$, $\theta_{12}$), where I have adopted
the standard parametrization \cite{Groom:2000in}
for the $3\times3$ MNSP \cite{Maki:1962mu,Pontecorvo:1968fh} matrix.
On the other hand, the CHOOZ 
result \cite{Apollonio:1999ae} tells us that
$|\theta_{13}|$ has to be small ($\sin^22\theta_{13}\lesssim 0.1$).
So the MNSP matrix looks like

\begin{eqnarray}
U_{MNSP} \simeq
\left(
\begin{array}{ccc}
c_\odot & s_\odot &  \epsilon\\
-s_\odot/\sqrt{2} &
c_\odot/\sqrt{2} & 1/\sqrt{2}\\
s_\odot/\sqrt{2} &
-c_\odot/\sqrt{2} & 1/\sqrt{2}\\
\end{array}
\right),\nonumber
\end{eqnarray}
where I have used $\theta_{23}\simeq \pi/4$,
$\sin^22\theta_{12}\equiv\sin^22\theta_\odot\simeq 0.8$
and $|\epsilon|\ll 1$.

The next thing to do is to determine $\theta_{13}$,
the sign of $\Delta m_{32}^2$ and the CP phase $\delta$.
During the past few years a lot of research have been done
on the possibilities of future long baseline experiments.
One is a super-beam experiment and the other one is
a neutrino factory.  The former is super intense conventional
neutrino beam which is obtained from pion decays while
the latter is from muon decays in a storage ring.
The background fraction $f_B$ in the case of
super-beams \cite{Itow:2001ee} is of order $10^{-2}$,
while in the case of a neutrino factory \cite{Cervera:2000vy}
it is or order $10^{-5}$.  The advantage of a neutrino factory
is such low background fraction and neutrino factories are
expected to enable us to determine $\theta_{13}$ and
the sign of $\Delta m_{32}^2$ ($\delta$) for
$\sin^22\theta_{13}\gtrsim 10^{-5}$ ($10^{-3}$), respectively.

In this talk I will mainly discuss measurements of
the CP phase $\delta$ at neutrino factories and will try
to clarify the reason why some group obtains different results
for the optimized muon energy and baseline.

\section{Measurements of the CP phase at neutrino factories}

Measurement of the CP phase in neutrino oscillations
is difficult not only because CP violating contribution
in the oscillation probability is in general small
but also because there is matter effect.
Namely, the dependence of the probabilities
for $\nu$ and
$\bar{\nu}$ on $\delta$ and $A$ is given by
$P(\nu_\mu\rightarrow\nu_e)=f(E,L;\theta_{ij},\Delta
m^2_{ij},\delta;A)$ and $P(\bar{\nu}_\mu\rightarrow\bar{\nu}_e)
=f(E,L;\theta_{ij},\Delta m^2_{ij},-\delta;-A)$, where $f$ is a
certain function and $A\equiv \sqrt{2}G_F N_e$ stands for the matter
effect.  The quantity obtained from experiments on $\bar{\nu}$ is not
$f(\cdots,-\delta;A)$, so that direct
comparison between $f(\cdots,\delta;A)$ and
$f(\cdots,-\delta;A)$ is impossible in a strict sense.
The situation here is different from the the $K^0-\bar{K}^0$ system
where the quantity
$N(K_L\rightarrow 2\pi)/N(K_L\rightarrow 3\pi)$
immediately gives us evidence for CP violation.
Hence I have to compromise and adopt a kind of indirect measurement
of CP violation, i.e., I deduce the values of
$\delta$ etc. by comparing the energy spectra of the data and of
the theoretical prediction with neutrino oscillations
assuming the three flavor mixing.
In determining $\delta$, there are other oscillation parameters as
well as the density of the Earth whose values are not exactly known,
so that I have to take into account correlations of errors of these
parameters.\footnote{Correlations of errors at neutrino factories
were studied in
Ref.~\cite{Cervera:2000kp,Bueno:2000fg,Koike:2002jf,Freund:2001ui,Pinney:2001xw}.}
Thus I introduce the following quantity
to see the significance of the case with nonvanishing $\delta$:
\begin{eqnarray}
\Delta\chi^2\equiv
\min_{{\ }_{\overline{\theta_{k\ell}},
\overline{\Delta m^2_{k\ell}},\overline{A}}}
&\displaystyle\sum_j &\left\{
{\left[
N_j(\nu_e\rightarrow\nu_\mu)-\bar{N}_j(\nu_e\rightarrow\nu_\mu)
\right]^2 \over \sigma^2_j}\right.\nonumber\\
&+& {\left[
N_j(\bar{\nu}_e\rightarrow\bar{\nu}_\mu)
-\bar{N}_j(\bar{\nu}_e\rightarrow\bar{\nu}_\mu)
\right]^2 \over \sigma^2_j}\nonumber\\
&+&{\left[
N_j(\nu_\mu\rightarrow\nu_\mu)-\bar{N}_j(\nu_\mu\rightarrow\nu_\mu)
\right]^2 \over \sigma^2_j}\nonumber\\
&+&\left. {\left[
N_j(\bar{\nu}_\mu\rightarrow\bar{\nu}_\mu)
-\bar{N}_j(\bar{\nu}_\mu\rightarrow\bar{\nu}_\mu)
\right]^2 \over \sigma^2_j}\right\},
\label{eqn:chi2}
\end{eqnarray}
where $N_j(\nu_\alpha\rightarrow\nu_\beta)
\equiv N_j(\nu_\alpha\rightarrow\nu_\beta;
\theta_{k\ell},\Delta m_{k\ell}^2,\delta,A)$,
$\bar{N}_j(\nu_\alpha\rightarrow\nu_\beta)
\equiv \bar{N}_j(\nu_\alpha\rightarrow\nu_\beta;
\overline{\theta_{k\ell}},\overline{\Delta m_{k\ell}^2},
\bar{\delta}=0,\bar{A})$ stand for the numbers of events
of the data and of the theoretical prediction with a vanishing
CP phase, respectively, and $\sigma^2_j$ stands for the error
which is given by the sum of the statistical and systematic
errors.  At neutrino factories appearance and disappearance
channels for $\nu$ and $\bar{\nu}$ are observed,
and I have included the numbers of events of all the
channels in (\ref{eqn:chi2}) to gain statistics.
In the present analysis, $N_j(\nu_\alpha\rightarrow\nu_\beta)$
is substituted by theoretical prediction with a CP phase $\delta$
and $\Delta\chi^2$ obviously vanishes if
$\delta=0$.\footnote{This is the reason why the quantity
in (\ref{eqn:chi2}) is denoted as $\Delta\chi^{2}$ instead of
absolute $\chi^2$.  $\Delta\chi^2$ represents
deviation from the best fit point rather than the goodness
of fit.}
The quantity $\Delta\chi^2$ reflects the strength of
the correlation of the parameters, i.e., if
$\Delta\chi^2$ turns out to be very small for
a certain value of $\bar{A}$ then the correlation
between $\delta$ and $A$ would be very strong and
in that case there would be no way to show
$\delta\ne0$.
To reject a hypothesis ``$\delta=0$'' at the 3$\sigma$
confidence level I demand
\begin{eqnarray}
\Delta\chi^2\ge\Delta\chi^2(3\sigma CL)
\label{eqn:3sigma}
\end{eqnarray}
where the right hand side stands for the value of
$\chi^2$ which gives the probability 99.7\%
in the $\chi^2$ distribution with a certain degrees freedom,
and $\Delta\chi^2(3\sigma CL)$=20.1 for 6 degrees freedom.
From (\ref{eqn:3sigma}) I get the condition for the detector size
to reject a hypothesis ``$\delta=0$'' at 3$\sigma$CL.

On the other hand, Koike et al.\cite{Koike:2002jf}
claim that another quantity
\begin{eqnarray}
\Delta{\tilde\chi}^2\equiv
\min_{{\ }_{\overline{\theta_{k\ell}},
\overline{\Delta m^2_{k\ell}},\overline{A}}}
\displaystyle\sum_j {1 \over \sigma^2_j}
\left[
{N_j(\nu_e\rightarrow\nu_\mu) \over
N_j(\bar{\nu}_e\rightarrow\bar{\nu}_\mu)}
-{\bar{N}_j(\nu_e\rightarrow\nu_\mu) \over
\bar{N}_j(\bar{\nu}_e\rightarrow\bar{\nu}_\mu)}
\right]^2
\label{eqn:chi2kos}
\end{eqnarray}
should be used since this particular combination improves
the correlation between $\delta$ and $A$.
In my opinion, however, both analyses with
$\Delta\chi^2$ and $\Delta{\tilde\chi}^2$ are based on
indirect measurements of CP violation anyway and there is no reason
why one has to discard other information on $\delta$.
In fact it turns out that (\ref{eqn:chi2kos}) is a combination
in which the correlation between  $\delta$ and $A$
improves for low energy and worsens at high energy.

It should be pointed out here that indirect measurements
of CP violation are also considered in the $B^0-\bar{B}^0$
system.\footnote{The term "indirect" in the B system is different
from that in neutrino oscillations.}
To measure the phase $\phi_1$, direct CP violating process
$B(\bar{B})\rightarrow J/\psi\, K_{s}$ is
used \cite{Carter:1981tk,Bigi:1981qs},
while those to measure $\phi_2$ and $\phi_3$ are
$B\rightarrow 2\pi$ \cite{Gronau:1990ka} and
$B\rightarrow D K$ \cite{Gronau:1991dp}, which are
not necessarily CP--odd processes.
Their strategy is to start with the three flavor framework,
to use the most effective process, which may or may not be
CP--odd, to determine the CP phases and to check unitarity
or consistency of the three flavor hypothesis.

\begin{figure}
\vglue 0.9cm
\begin{center}
\begin{picture}(10,10)(135,170)
\put(-15,-15){\makebox(100,100){\Large{$\phi_{{}_3}$}}}
\put(5, 100){\makebox(100,100){\Large{$\phi_{{}_2}$}}}
\put(130,-15){\makebox(100,100){\Large{$\phi_{{}_1}$}}}
\put(30,180){\makebox(100,100){\Large{$\phi_{{}_2}$: from $B\rightarrow 2\pi$}}}
\put(0,-90){\makebox(100,100){\Large{$\phi_{{}_3}$: from $B\rightarrow D K$}}}
\put(140,100){\makebox(100,100){\Large{$\phi_{{}_1}$: from $B(\bar{B})\rightarrow J/\psi\, K_{s}$}}}
\put(110,75){\makebox(100,100){\Large{(direct)}}}
\put(0,0){\line(1,0){240}}
\put(0,0){\line(1,5){40}}
\put(40,200){\line(1,-1){200}}
\bezier{150}(4,20)(20,17)(20,0)
\bezier{150}(220,0)(220,9)(226,14)
\bezier{150}(54,186)(47,179)(36,181)
\end{picture}
\end{center}
\vglue 7.9cm
\caption{Unitarity triangle in the B meson system.
The phase $\phi_{1}$ is measured by a direct CP violating process
$B(\bar{B})\rightarrow J/\psi\, K_{s}$,
whereas the most promising way
to determine $\phi_{2}$ and $\phi_{3}$ is through
$B\rightarrow 2\pi$ and
$B\rightarrow D K$,
which are not direct CP violating processes.
}
\label{fig:Fig1}
\end{figure}
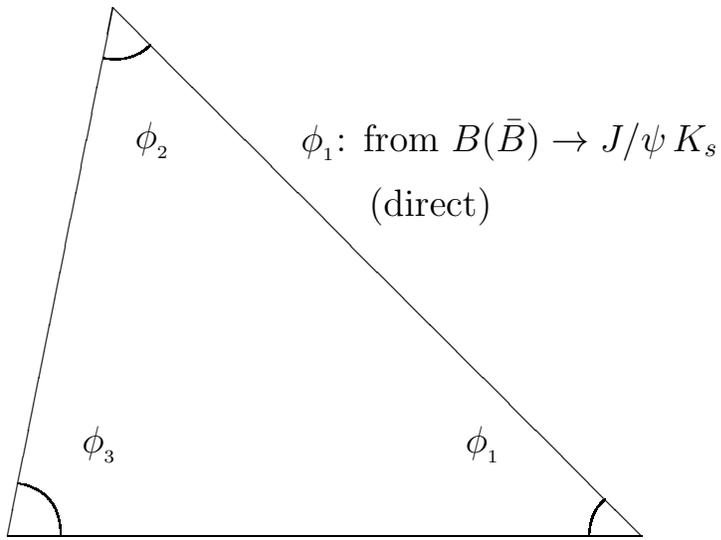

%\newpage
%\hglue -5cm
\begin{table}
%\rotate[r]{
\vglue 0.9cm
%\vglue 10cm
\hglue -1.3cm
\begin{tabular}{|p{8mm}|p{17mm}|p{10mm}|p{6mm}|p{23mm}|p{12mm}|p{14mm}|p{18mm}|}
\hline
{Ref.}&
{correlations of $\theta_{ij}$,$m_{ij}^2$}&
{$|\Delta A/A|$}&
{$f_B$}&
{$\Delta m_{21}^2/10^{-5}$eV$^2$}&
{$E_{th}$/GeV}&
{optimized $E_{\mu}$/GeV}&
{optimized $L$/km}\\
\hline
\hline
KOS \cite{Koike:2002jf}&
included&
10\%&
0&
5&
1&
$\lesssim$6&
600 -- 800\\
\cline{5-8}
&&&&10&1&
$\lesssim$50&
500 -- 2000\\
\hline
FHL \cite{Freund:2001ui}&
included&
0&
0&
10&
4&
30 -- 50&
2800 -- 4500\\
\hline
PY \cite{Pinney:2001xw}&
included&
5\%&
$10^{-5}$&
3.2&
0.1&
$\sim$50&
$\sim$3000\\
\cline{4-4}\cline{7-8}
&&&$10^{-3}$&
&&$\sim$20&
$\sim$1000\\
\cline{3-4}\cline{7-8}
&&10\%&
$10^{-5}$&
&&$\sim$15&
$\sim$1000\\
\cline{4-4}\cline{7-8}
&&&$10^{-3}$&
&&$\sim$10&
$\sim$800\\
\cline{3-4}\cline{7-8}
&&20\%&
$10^{-5}$&
&&$\sim$8&
$\sim$500\\
\cline{4-4}\cline{7-8}
&&&$10^{-3}$&
&&$\sim$6&
$\sim$500\\
\hline
H \cite{huber}&
included&
10\%&
0&
10&
0.1&
$\gtrsim$20&
$\gtrsim$2000\\
\cline{5-5}\cline{7-8}
&&&&3.5&&
$\sim$25&
$\sim$1500\\
\hline
\end{tabular}
\caption{Comparison of different works
% \cite{Carter:1981tk,Bigi:1981qs}.
The reference value is
$\sin^{2}2\theta_{13}=0.1$, $10^{21} \mu$s.}
%}
\label{tbl:table1}
\end{table}

%\newpage

\begin{figure}
\vglue -2.80cm \hglue -1.0cm
\epsfig{file=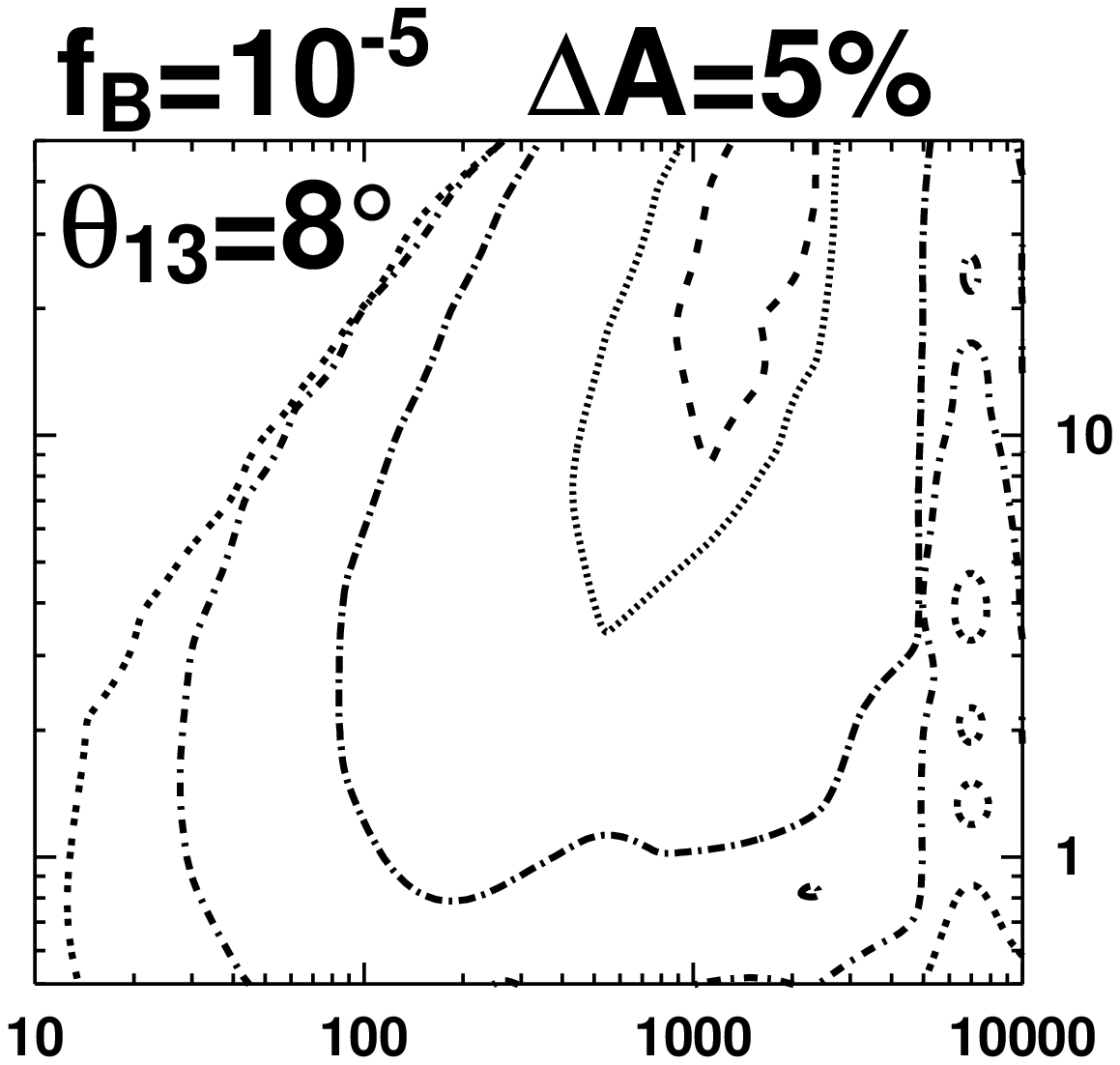,width=7cm}
\vglue -7.9cm \hglue 3.8cm
\epsfig{file=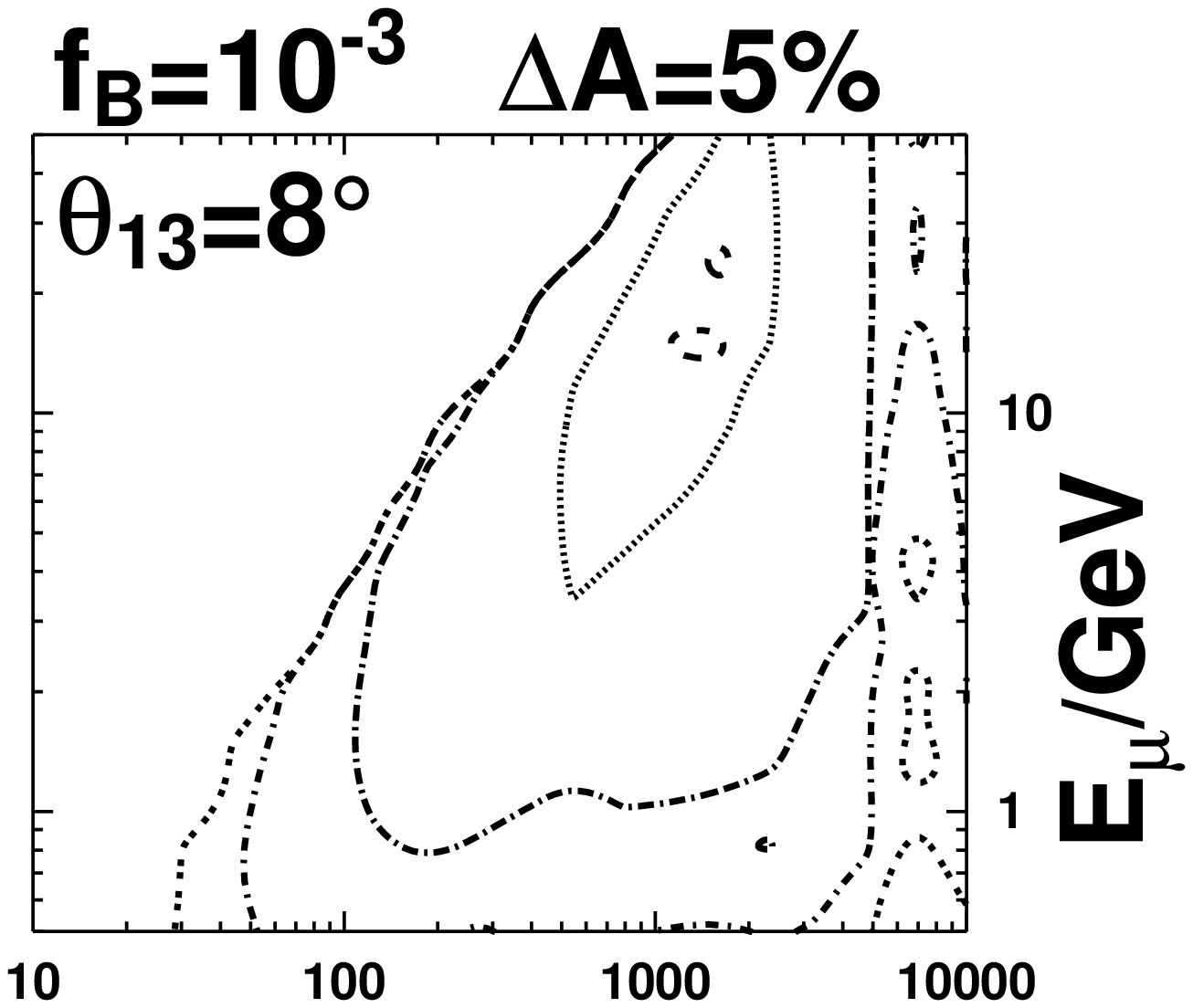,width=7cm}
\vglue -3.5cm \hglue -1.0cm
\epsfig{file=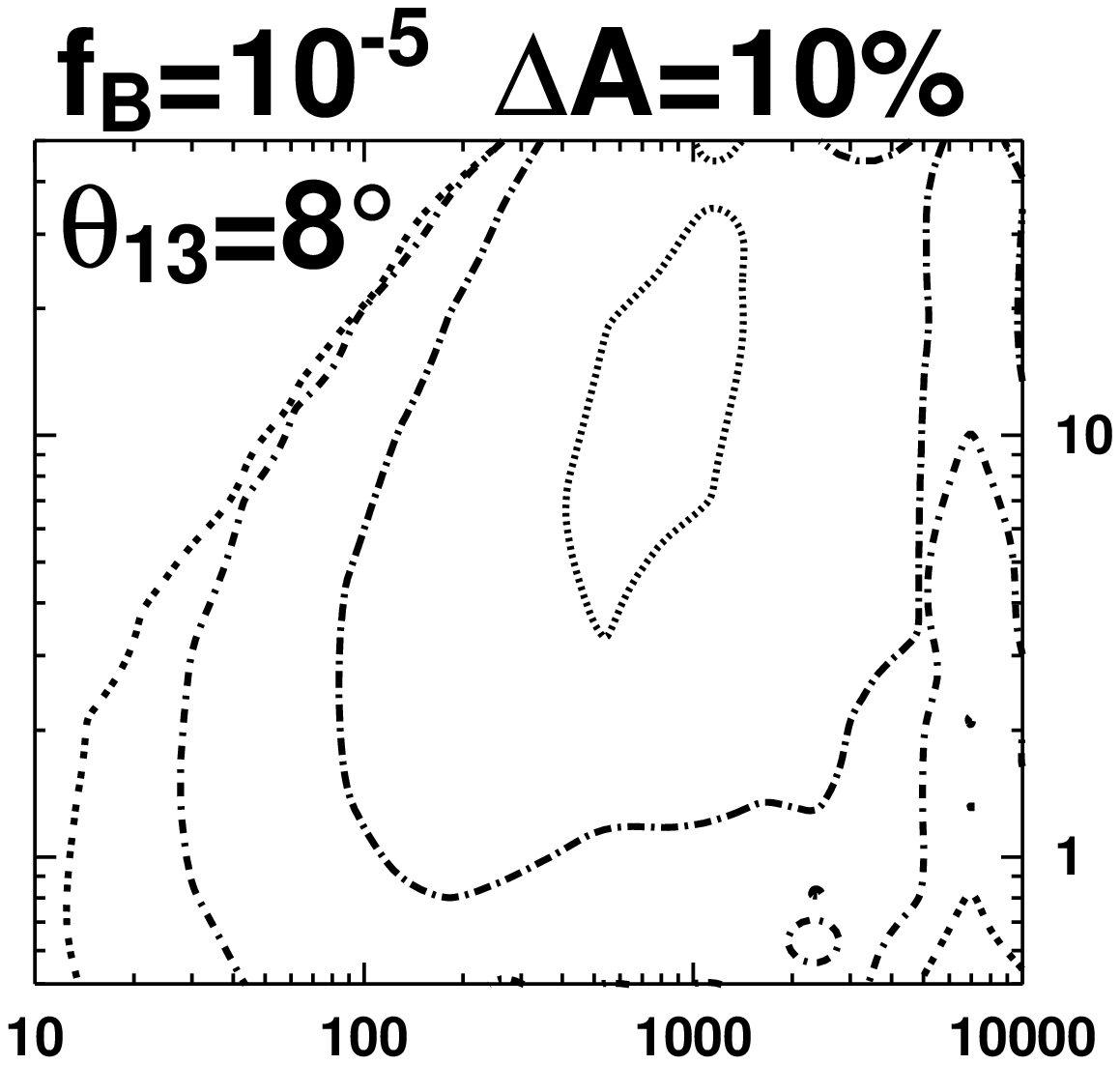,width=7cm}
\vglue -7.9cm \hglue 3.8cm
\epsfig{file=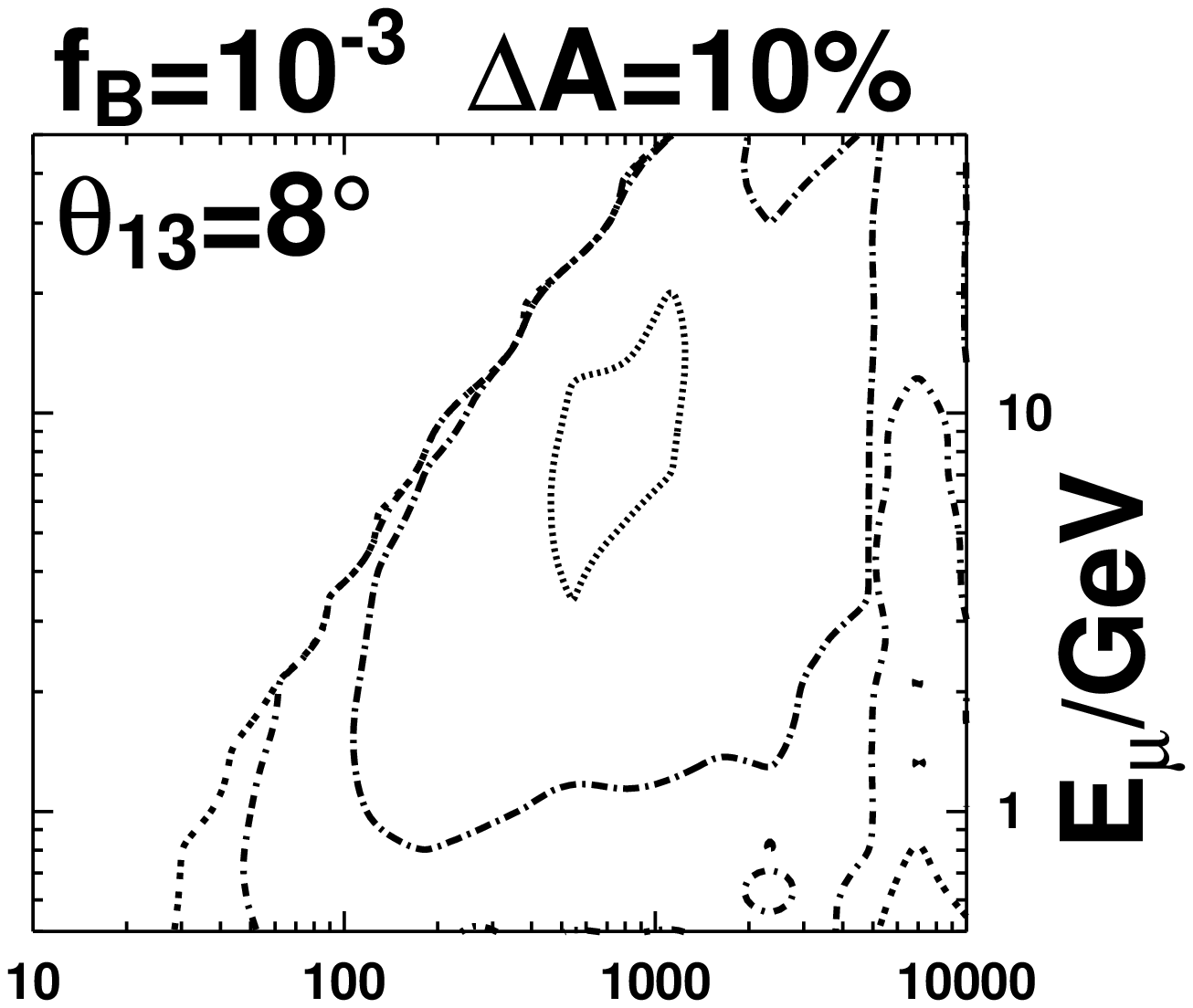,width=7cm}
\vglue -3.5cm \hglue -1.0cm
\epsfig{file=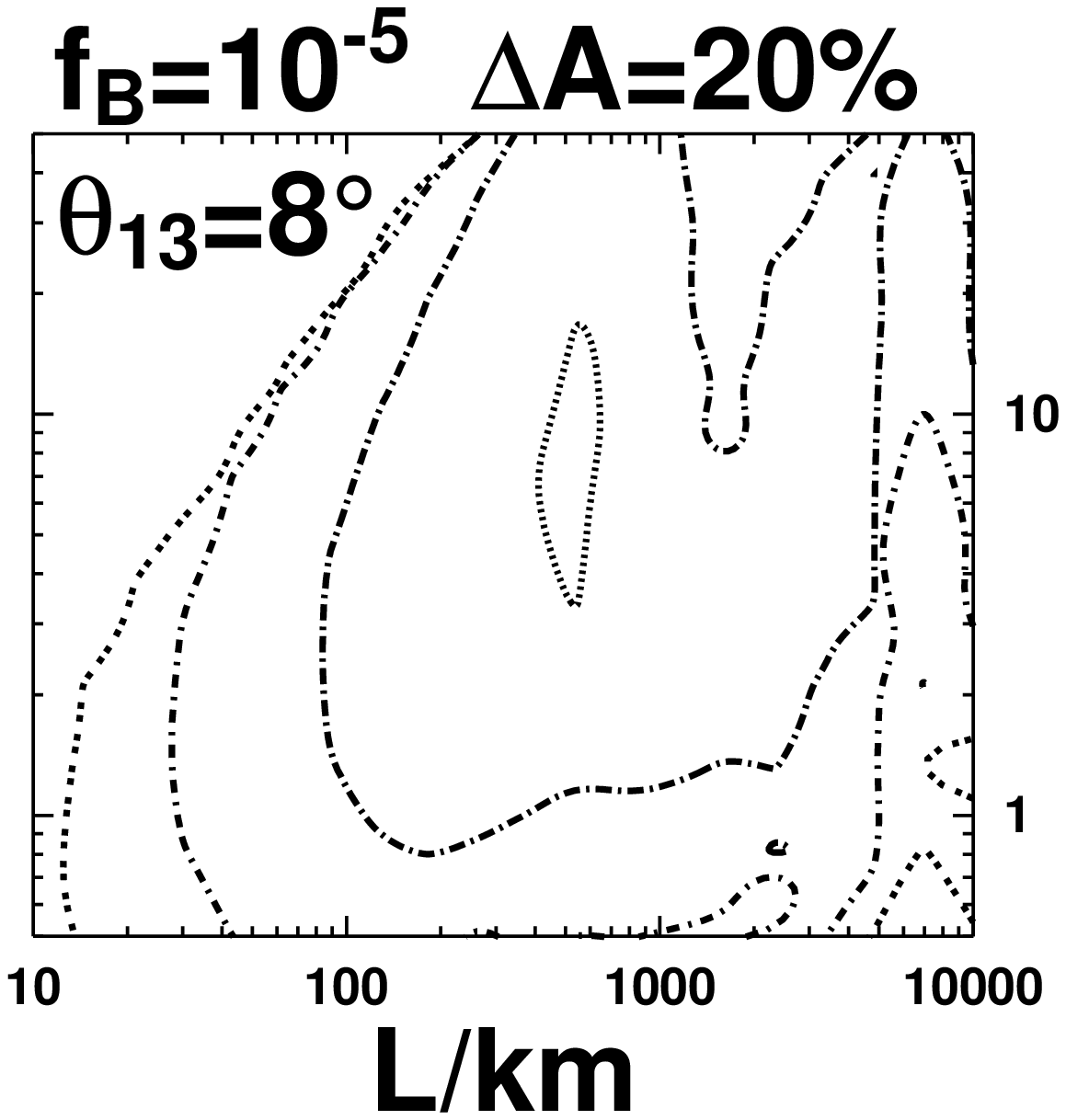,width=7cm}
\vglue -7.9cm \hglue 3.8cm
\epsfig{file=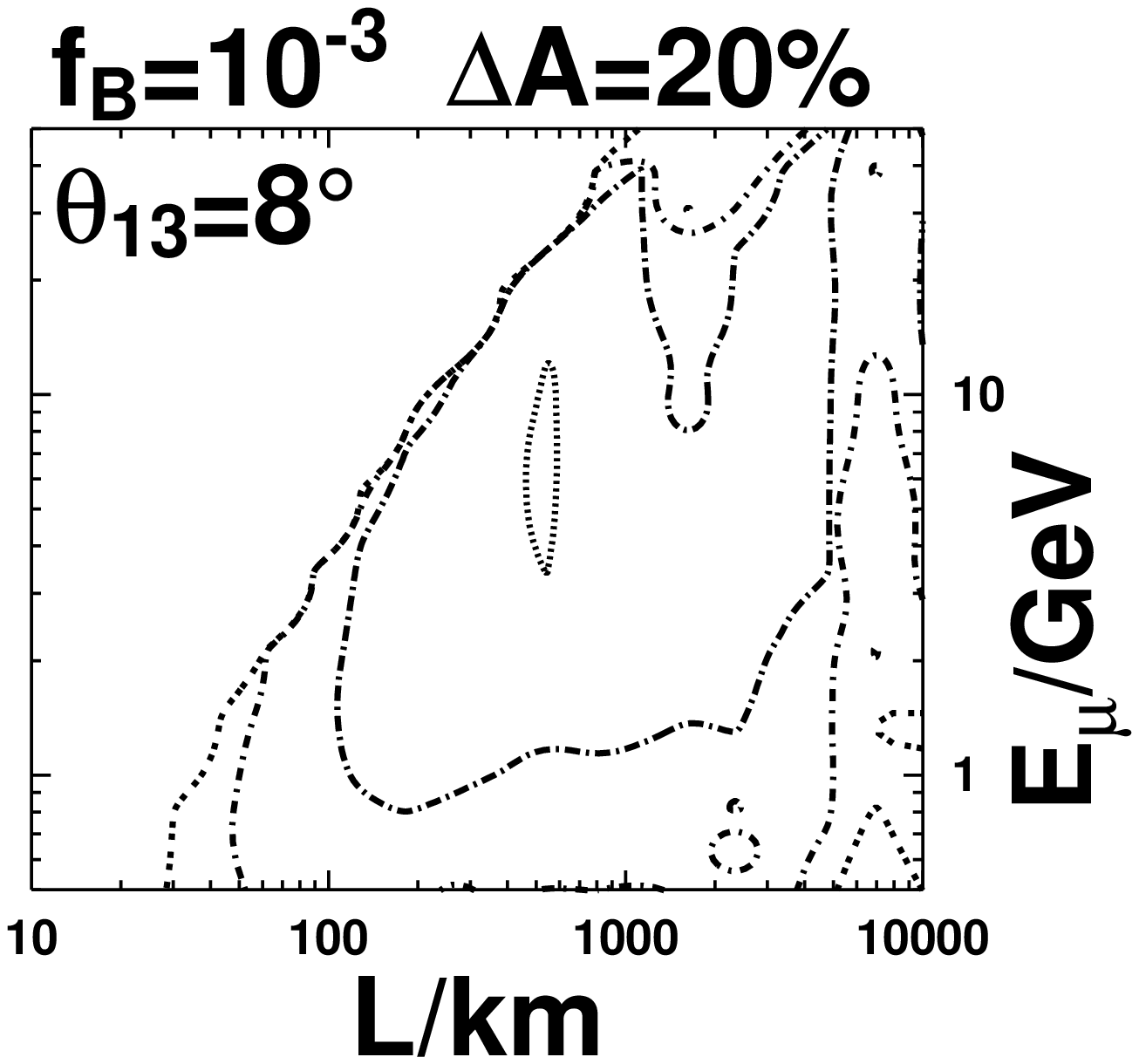,width=7cm}
\vglue -4.0cm\hglue -10.3cm
\epsfig{file=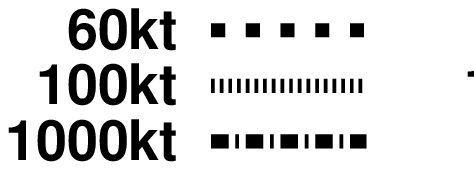,width=14cm}
\vglue -12.0cm \hglue 3.3cm
\caption{\small
The contour plot of equi-number of data size required
to reject a hypothesis $\bar{\delta}=0$ at $3\sigma$
with
the background fraction $f_B=10^{-5}$ or $10^{-3}$
and the uncertainty of the matter effect $\Delta A$=5\%, 10\% or 20\%.
}
\label{fig:Fig2}
\vglue -0.5cm 
\end{figure}

\begin{figure}
\begin{center}
\vglue -5cm \hglue -3cm
\epsfig{file=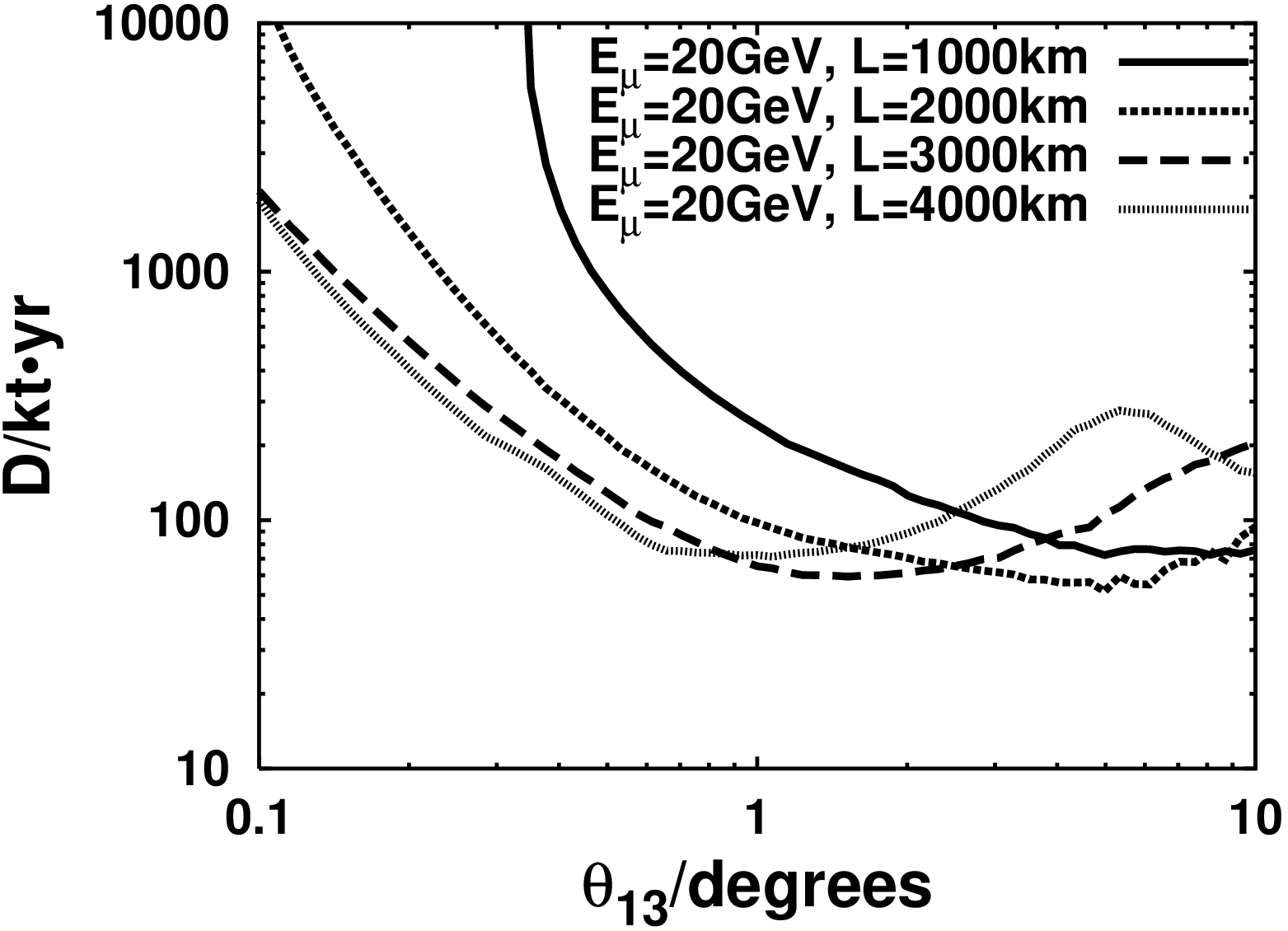,width=6cm}
\vglue -2.5cm \hglue -3cm
\epsfig{file=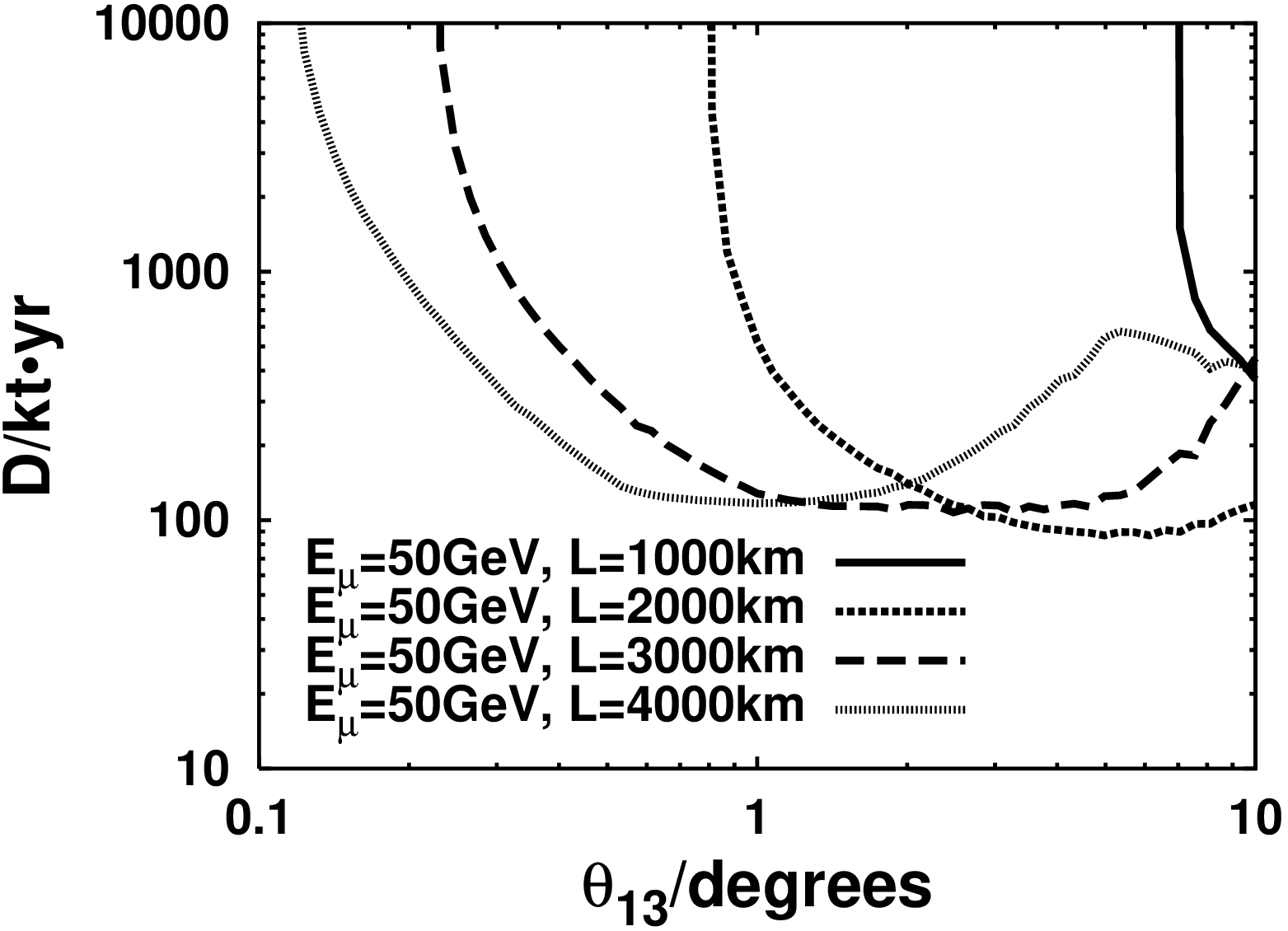,width=6cm}
\vglue 1.0cm
\caption{Data size (kt$\cdot$yr) required to reject a hypothesis of
$\delta=0$ at 3$\sigma$ when the true value is $\delta=\pi/2$,
in the case of a neutrino factory with $10^{21}$ useful muon decays
per year and a background fraction $f_B=10^{-3}$.}
\end{center}
\label{fig:jp}
\end{figure}

The optimized muon energy $E_\mu$ and baseline $L$
for the measurement of CP the phase at a neutrino
factory have been investigated by several groups
by taking into consideration the correlations of
$\delta$ and all other parameters and the results
are summarized in Table \ref{tbl:table1}.
The results in Ref.\cite{Pinney:2001xw} are given in Fig. \ref{fig:Fig2},
which shows that the more uncertainty I have in the density,
the shorter baseline I have to choose because
the correlation between $\delta$ and $A$
becomes stronger for larger baseline and muon energy.
The works \cite{Freund:2001ui,Pinney:2001xw,huber},
which basically used $\Delta\chi^2$ in (\ref{eqn:chi2}),
agree with each other to certain extent (there are some differences
on the reference values for the oscillation parameters)
while the result by Koike et al.\cite{Koike:2002jf}
is quite different from others.  This discrepancy
is due to the fact that they adopted $\Delta{\tilde\chi}^2$ in
(\ref{eqn:chi2kos}) as was mentioned above.
There are slight differences between the results by
the group \cite{Freund:2001ui,huber} and those by
the other \cite{Pinney:2001xw} and this appears to
come from different statistical treatments.
In the future it should be studied what makes a difference
to get the optimum set $(E_\mu, L)$.
The detector size required to reject a hypothesis "$\delta$=0" as
a function of $\theta_{13}$ is given in
Figure \ref{fig:jp} (taken from Ref.\cite{Pinney:2001bj}).
Figure \ref{fig:jp}
shows the sensitivity of neutrino factories to the CP phase.

\section{High energy behaviors of $\Delta\chi^2$}

Some people have questioned whether the sensitivity to the CP phase at
a neutrino factory increases infinitely as the muon energy increases,
and Lipari\cite{Lipari:2001ds} concluded that the sensitivity is lost
at high energy.  In the work \cite{Pinney:2001xw} the behaviors
of $\Delta\chi^2$ in (\ref{eqn:chi2}) was studied analytically
for high muon energy and it was shown
after the correlations between $\delta$ and any other oscillation
parameter or $A$ is taken into account that
\begin{eqnarray}
\Delta \chi^2
\propto \left({J \over \sin\delta}\right)^2
{1 \over E_\mu}\left(
\sin\delta+\mbox{\rm const}{\Delta m^2_{32}L \over E_\mu}
\cos\delta\right)^2
\label{eqn:largeemu}
\end{eqnarray}
for large $E_\mu$, where
$J\equiv (c_{13} / 8)\sin2\theta_{12}\sin2\theta_{13}
\sin2\theta_{23}\sin\delta$ stands for the Jarlskog parameter.
The behavior (\ref{eqn:largeemu}) is the same as that for
$\Delta{\tilde\chi}^2$ in (\ref{eqn:chi2kos}) and it is
qualitatively consistent with the claim by Lipari \cite{Lipari:2001ds}.
It is remarkable that (\ref{eqn:largeemu}) is different from
a naively expected behavior
\begin{eqnarray}
\Delta\chi^2_{\mbox{\rm{\scriptsize  naive}}}\propto E_\mu(\cos\delta-1)^2.
\nonumber
\end{eqnarray}
This is because the correlation between $\delta$ and
other parameters is taken into account.
Koike et al.\cite{Koike:2002jf} criticize the analysis with
$\Delta\chi^2$ by saying that this quantity looks
mainly at the CP conserving part at high muon energy
(10GeV$\lesssim E_\mu\lesssim$50GeV).
The behavior (\ref{eqn:largeemu}) indicates, however,
that CP violating part becomes dominant
{\it after} the correlation between $\delta$ and
other parameters is taken into consideration and
the criticism by Koike et al.\cite{Koike:2002jf} does
not apply.

\section{Parameter degeneracy}

\begin{figure}[ht]
\vglue -0.2cm 
\centerline{\protect\hbox{
\psfig{file=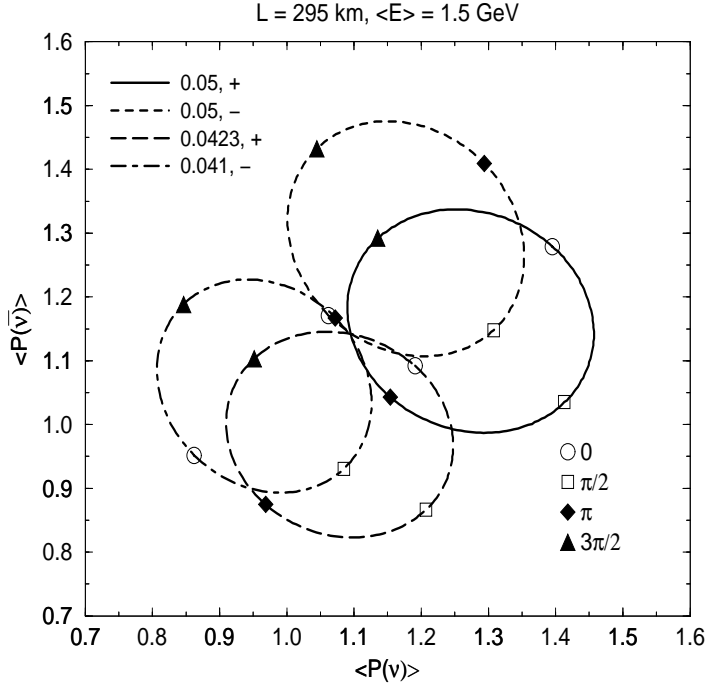,height=10.2cm,width=10.8cm}
}}
\vglue -1.0cm 
\caption{\small
Illustration of the clover-leaf ambiguity in terms of CP 
trajectory diagram.  Four solutions exist for given values of 
$P(\nu_\mu\to\nu_e)$ and
$P(\bar{\nu}_\mu\to\bar{\nu}_e)$.
}
\label{fig:Fig3}
\vglue 1.0cm 
\end{figure}

The discussions in the previous sections
have been focused on rejection of
a hypothesis ``$\delta=0$''.  Once
$\delta$ is found to be nonvanishing, it becomes important
to determine the precise value of $\delta$.
It has been pointed out that various kinds of parameter degeneracy
exist.  Burguet-Castell et al.\cite{Burguet-Castell:2001ez}
found degeneracy in ($\delta$, $\theta_{13}$),
Minakata and Nunokawa \cite{Minakata:2001qm} found
the one in the sign of $\Delta m^2_{32}$, and
Barger et al.\cite{Barger:2001yr} found the one
in the sign of $\pi/4-\theta_{23}$.
To understand the four-fold ambiguity it is instructive
to draw a trajectory in the $P(\nu_\mu\rightarrow\nu_e)$ --
$P(\bar{\nu}_\mu\rightarrow\bar{\nu}_e)$ plane
(See Figure \ref{fig:Fig3} taken from Ref.\cite{Minakata:2001rk}).
From Figure \ref{fig:Fig3} it is obvious that
there exists four-fold ambiguity for a given set of the oscillation parameters.
Also the position of the ellipse
has degeneracy in interchanging
$\theta_{23}\leftrightarrow\pi/2-\theta_{23}$, so that
in general eight-fold degeneracy is expected. \cite{Barger:2001yr}
It was proposed \cite{Burguet-Castell:2001ez,Minakata:2001qm,Barger:2001yr}
to do experiments at two baselines to remove this degeneracy.
In reality one has to evaluate numbers of events
for $\nu_\mu\rightarrow\nu_e$ and $\bar{\nu}_\mu\rightarrow\bar{\nu}_e$
(Notice that Figure \ref{fig:Fig3} deals with the probabilities only),
and the correlations of errors have to be taken into
consideration as well, so determination of the oscillation
parameters is even more difficult than what Figure \ref{fig:Fig3} indicates.

\section{Summary}
People in the field reached consensus that
neutrino factories can measure the CP phase $\delta$ with
the detector size larger than $10^{21}\mu\cdot$100kt$\cdot$yr
for $\sin^22\theta_{13}\gtrsim 10^{-3}$ unless
$|\delta|$ is small.  The detailed study
on the optimized muon energy and baseline still needs
to be done.

\section*{Acknowledgments}
I would like to thank Noriaki Kitazawa for useful discussions on
CP violation in the B meson system and for providing references
on the subject.  This
research was supported in part by a Grant-in-Aid for Scientific
Research of the Ministry of Education, Science and Culture,
\#12047222, \#13640295.


\newpage
\section*{References}
\begin{thebibliography}{99}

% \cite{Kajita:2001mr}
\bibitem{Kajita:2001mr}
T.~Kajita and Y.~Totsuka,
%``Observation of atmospheric neutrinos,''
Rev.\ Mod.\ Phys.\  {\bf 73}, 85 (2001).

% \cite{Bahcall:2000kh}
\bibitem{Bahcall:2000kh}
J.~N.~Bahcall,
%``Solar neutrinos: An overview,''
Phys.\ Rept.\  {\bf 333}, 47 (2000).
%%CITATION = PRPLC,333,47;%%

% \cite{Groom:2000in}
\bibitem{Groom:2000in}
D.~E.~Groom {\it et al.}  [Particle Data Group Collaboration],
%``Review of particle physics,''
Eur.\ Phys.\ J.\ C {\bf 15}, 1 (2000).
%%CITATION = EPHJA,C15,1;%%

% \cite{Apollonio:1999ae}
\bibitem{Apollonio:1999ae}
M.~Apollonio {\it et al.}  [CHOOZ Collaboration],
%``Limits on neutrino oscillations from the CHOOZ experiment,''
Phys.\ Lett.\ B {\bf 466}, 415 (1999)
[arXiv:hep-ex/9907037].
%%CITATION = HEP-EX 9907037;%%

% \cite{Maki:1962mu}
\bibitem{Maki:1962mu}
Z.~Maki, M.~Nakagawa and S.~Sakata,
%``Remarks On The Unified Model Of Elementary Particles,''
Prog.\ Theor.\ Phys.\  {\bf 28} (1962) 870.
%%CITATION = PTPKA,28,870;%%

% \cite{Pontecorvo:1968fh}
\bibitem{Pontecorvo:1968fh}
B.~Pontecorvo,
%``Neutrino Experiments And The Question Of Leptonic-Charge  Conservation,''
Sov.\ Phys.\ JETP {\bf 26} (1968) 984
[Zh.\ Eksp.\ Teor.\ Fiz.\  {\bf 53} (1968) 1717].
%%CITATION = SPHJA,26,984;%%

% \cite{Itow:2001ee}
\bibitem{Itow:2001ee}
Y.~Itow {\it et al.},
%``The JHF-Kamioka neutrino project,''
arXiv:hep-ex/0106019.
%%CITATION = HEP-EX 0106019;%%

% \cite{Cervera:2000vy}
\bibitem{Cervera:2000vy}
A.~Cervera, F.~Dydak and J.~Gomez Cadenas,
%``A large magnetic detector for the neutrino factory,''
Nucl.\ Instrum.\ Meth.\ A {\bf 451}, 123 (2000).
%%CITATION = NUIMA,A451,123;%%

% \cite{Cervera:2000kp}
\bibitem{Cervera:2000kp}
A.~Cervera, A.~Donini, M.~B.~Gavela, J.~J.~Gomez Cadenas, P.~Hernandez, O.~Mena and S.~Rigolin,
%``Golden measurements at a neutrino factory,''
Nucl.\ Phys.\ B {\bf 579}, 17 (2000)
[Erratum-ibid.\ B {\bf 593}, 731 (2000)]
[arXiv:hep-ph/0002108].
%%CITATION = HEP-PH 0002108;%%

% \cite{Bueno:2000fg}   
\bibitem{Bueno:2000fg}
A.~Bueno, M.~Campanelli and A.~Rubbia,
%``Physics potential at a neutrino factory: Can we benefit from more than  just detecting muons?,''
Nucl.\ Phys.\ B {\bf 589}, 577 (2000)
[arXiv:hep-ph/0005007].
%%CITATION = HEP-PH 0005007;%%

% \cite{Koike:2002jf}
\bibitem{Koike:2002jf}
M.~Koike, T.~Ota and J.~Sato,
%``Ambiguities of theoretical parameters and CP/T violation in neutrino  factories,''
Phys.\ Rev.\ D {\bf 65}, 053015 (2002)
[arXiv:hep-ph/0011387].
%%CITATION = HEP-PH 0011387;%%

% \cite{Freund:2001ui}
\bibitem{Freund:2001ui}
M.~Freund, P.~Huber and M.~Lindner,
%``Systematic exploration of the neutrino factory parameter space  including errors and correlations,''
Nucl.\ Phys.\ B {\bf 615}, 331 (2001)
[arXiv:hep-ph/0105071].
%%CITATION = HEP-PH 0105071;%%

% \cite{Pinney:2001xw}
\bibitem{Pinney:2001xw}
J.~Pinney and O.~Yasuda,
%``Correlations of errors in measurements of CP violation at neutrino  factories,''
Phys.\ Rev.\ D {\bf 64}, 093008 (2001)
[arXiv:hep-ph/0105087].
%%CITATION = HEP-PH 0105087;%%

% \cite{Carter:1981tk}
\bibitem{Carter:1981tk}
A.~B.~Carter and A.~I.~Sanda,
%``CP Violation In B Meson Decays,''
Phys.\ Rev.\ D {\bf 23}, 1567 (1981).
%%CITATION = PHRVA,D23,1567;%%

% \cite{Bigi:1981qs}
\bibitem{Bigi:1981qs}
I.~I.~Bigi and A.~I.~Sanda,
%``Notes On The Observability Of CP Violations In B Decays,''
Nucl.\ Phys.\ B {\bf 193}, 85 (1981).
%%CITATION = NUPHA,B193,85;%%

% \cite{Gronau:1990ka}
\bibitem{Gronau:1990ka}
M.~Gronau and D.~London,
%``Isospin Analysis Of CP Asymmetries In B Decays,''
Phys.\ Rev.\ Lett.\  {\bf 65}, 3381 (1990).
%%CITATION = PRLTA,65,3381;%%

% \cite{Gronau:1991dp}
\bibitem{Gronau:1991dp}
M.~Gronau and D.~Wyler,
%``On determining a weak phase from CP asymmetries in charged B decays,''
Phys.\ Lett.\ B {\bf 265}, 172 (1991).
%%CITATION = PHLTA,B265,172;%%

\bibitem{huber}
P.~Huber, private communication.

% \cite{Pinney:2001bj}
\bibitem{Pinney:2001bj}
J.~Pinney,
%``Correlations of errors in a CP-violation neutrino factory experiment,''
[arXiv:hep-ph/0106210].
%%CITATION = HEP-PH 0106210;%%

% \cite{Lipari:2001ds}
\bibitem{Lipari:2001ds}
P.~Lipari,
%``CP violation effects and high energy neutrinos,''
Phys.\ Rev.\ D {\bf 64}, 033002 (2001)
[arXiv:hep-ph/0102046].
%%CITATION = HEP-PH 0102046;%%

% \cite{Burguet-Castell:2001ez}
\bibitem{Burguet-Castell:2001ez}
J.~Burguet-Castell, M.~B.~Gavela, J.~J.~Gomez-Cadenas, P.~Hernandez and O.~Mena,
%``On the measurement of leptonic CP violation,''
Nucl.\ Phys.\ B {\bf 608}, 301 (2001)
[arXiv:hep-ph/0103258].
%%CITATION = HEP-PH 0103258;%%

% \cite{Minakata:2001qm}
\bibitem{Minakata:2001qm}
H.~Minakata and H.~Nunokawa,
%``Exploring neutrino mixing with low energy superbeams,''
JHEP {\bf 0110}, 001 (2001)
[arXiv:hep-ph/0108085].
%%CITATION = HEP-PH 0108085;%%

% \cite{Barger:2001yr}
\bibitem{Barger:2001yr}
V.~Barger, D.~Marfatia and K.~Whisnant,
%``Breaking eight-fold degeneracies in neutrino CP violation, mixing, and  mass hierarchy,''
arXiv:hep-ph/0112119.
%%CITATION = HEP-PH 0112119;%%

% \cite{Minakata:2001rk}
\bibitem{Minakata:2001rk}
H.~Minakata and H.~Nunokawa,
%``CERN to Gran Sasso: An ideal distance for superbeam?,''
arXiv:hep-ph/0111131.
%%CITATION = HEP-PH 0111131;%%

\end{thebibliography}
\end{document}